\documentclass{nature}
\def\Vbg{V_{\mathrm{BG}}}

\def\dEf{\delta E_{F}}

\def\rhoxx{\rho_{\mathrm{xx}}}
\def\rhoxy{\rho_{\mathrm{xy}}}
\def\sigmaxy{\sigma_{\mathrm{xy}}}
\def\Bz{B_{\perp}}
\def\Bt{B_{\mathrm{tot}}}
\def\Ec{E_{\mathrm{c}}}
\def\EZ{E_{\mathrm{Z}}}
\def\lB{l_{\mathrm{B}}}
\def\={\,=\,}
\def\kB{k_{\mathrm{B}}}
\def\Pm{\,\pm\,}

\usepackage{amssymb}
\usepackage{amsmath}
\usepackage{graphicx}
\usepackage{graphics}
\usepackage{color}
\usepackage[square,sort,comma,numbers,super,sort&compress]{natbib}
\usepackage{multirow}
\usepackage{upgreek}
\usepackage[applemac]{inputenc}
\usepackage{cite}
\usepackage{upgreek}

\title{Composite Fermions and Broken Symmetries in Graphene}

\author{F. Amet$^{1,\dagger}$, A. J. Bestwick$^{2}$, J. R. Williams$^{2}$, L. Balicas$^3$, K. Watanabe$^4$, T. Taniguchi$^{4}$ \& D. Goldhaber-Gordon$^{1,2,\dagger}$}

\begin{document}

\maketitle

\begin{affiliations}
 \item Department of Applied Physics, Stanford University, CA 94305 Stanford, USA.
 \item Department of Physics, Stanford University, CA 94305 Stanford, USA.
 \item National High Magnetic Field Laboratory, FL 32310-3706 Tallahassee, USA.
 \item Advanced Materials Laboratory, National Institute for Materials Science, 1-1 Namiki, Tsukuba, 305-0044, Japan.
 \\$\dagger$ Contact: goldhaber-gordon@stanford.edu
\end{affiliations}

\begin{abstract}
The electronic properties of graphene are described by a Dirac Hamiltonian with a fourfold symmetry of spin and valley. This symmetry may yield novel fractional quantum Hall (FQH) states at high magnetic field depending on the relative strength of symmetry breaking interactions. However, observing such states in transport remains challenging in graphene, as they are easily destroyed by disorder. In this work, we observe in the first two Landau levels ($\nu\leq6$) the composite-fermion sequences of FQH states at $p/(2p ± 1)$ between each integer filling factor. In particular, odd numerator fractions appear between $\nu$$\,=\,$1 and $\nu$$\,=\,$2, suggesting a broken valley symmetry, consistent with our observation of a gap at charge neutrality and zero field. Contrary to our expectations, the evolution of gaps in a parallel magnetic field suggests that states in the first Landau level are not spin-polarized even up to very large out of plane fields.
\end{abstract}

\paragraph{Introduction}
$\\$
In a large magnetic field $B$, the band structure of two-dimensional electrons becomes a discrete set of highly degenerate Landau levels (LLs)~\cite{Zhang2005,Geim2007,Novoselov2005,Neto2009}. With kinetic energy quenched, electron interactions determine the ground state of a partially filled LL. This yields new incompressible phases known as fractional quantum Hall (FQH) states~\cite{Tsui1982, Laughlin1983}, where the longitudinal resistance $\rhoxx$ vanishes exponentially at low temperatures and the transverse resistance $\rhoxy$ is quantized as $h/\nu e^{2}$, where $e$ is the electron charge, $h$ Planck's constant, and $\nu$$\,=\,$$nh/eB$ the number of filled LLs for an electron density $n$.  

In GaAs quantum wells, the most robust FQH states are observed when $\nu$$\,=\,$$p/(2kp\pm1)$$\,=\,$1/3, 2/3, 2/5, 3/5...($k$ and $p$ integers). This may be understood within the composite fermion (CF) theory, where each electron is imagined to bind an even number 2k of magnetic flux quanta $\phi_{\mathrm{0}}$$\,=\,$$\frac{h}{e}$~\cite{Halperin1983,Jain1989,Kivelson,Girvin}. CFs experience an effective residual magnetic field $B^{*}$$\,=\,$$B-2kn\phi_{\mathrm{0}}$, and incompressible phases occur when the number of filled CF Landau levels $p$$\,=\,$$n\phi_{0}/B^{*}$ is integer.

In graphene, the ground state in each of these FQH phases is characterized by the two internal degrees of freedom of the Hamiltonian, the spin and the valley isospin, the latter originating from the hexagonal crystal structure of graphene. These grant Landau levels an approximate SU(4) symmetry~\cite{Geim2007,Neto2009,Zhang2005, Novoselov2005}, broken at high magnetic fields~\cite{Zhang2006} by Zeeman splitting $\EZ$$\,=\,$$g\mu_{\mathrm{B}}B$ and valley symmetry breaking on order $(a/\lB)\Ec\sim ae^{2}/\epsilon\lB^2$~\cite{Alicea2007,Young2012,Goerbig2006}, where $a$ is the graphene lattice constant, $\lB$ the magnetic length, $\Ec$ the typical energy of Coulomb interactions at a length scale $\lB$, and $\epsilon$ the dielectric constant. While the nature of the broken-symmetry phases at integer filling factors has been under intense scrutiny~\cite{Zhang2006,Alicea2007,Young2012,Goerbig2006,Kharitonov2012,Nomura2006}, little is known about the impact of symmetry-breaking interactions on FQH states~\cite{Goerbig2007,Toke2007,Khveshchenko2007,Shibata2009,Papic2009,Papic2010,Toke2012, Peterson2014}. %In fact, the absence of certain FQH states in the n=0 Landau level, such as $\nu\,=\,5/3$, was attributed to the incomplete breaking of the SU(4) symmetry

\paragraph{Results}\subparagraph{Characterization at low magnetic field}
$\\$
Observation of FQH states requires that disorder-induced Fermi level fluctuations $\dEf$ be smaller than FQH energy gaps. To minimize $\dEf$, we place monolayer graphene on an atomically-flat flake of hexagonal boron nitride~\cite{Dean2009,Xue2011}which is typically 15 to 25$\,$nm thick in the seven heterostructures we studied. The carrier density is tuned with a voltage $\Vbg$ applied to a graphite back-gate [Fig. 1a], which also acts as a screening layer, recently shown to make the potential landscape in graphene devices cleaner~\cite{Ponomarenko2012,Amet2013,Hunt2013}. Potential fluctuations should be suppressed on length scales larger than the distance to the back gate, while electron interactions remain on the scale of the magnetic length $\lB$$\,=\,$26$\,$nm$/\sqrt{B[\mathrm{T}]}$, less than the distance to the back gate for the relevant magnetic field range. The devices included in this study have field-effect mobilities ranging from 4$\times 10^{5}$ to $10^{6}$ cm$^{2}$/Vs, as extracted from a linear fit to the conductivity $\sigma(n)$ at low density (typically $n$$\,<\,$2$\times10^{11}$ cm$^{-2}$). This high quality is seen from the longitudinal resistivity $\rhoxx$, plotted in logarithmic scale as a function of the carrier density for Device A, at temperature $T$$\,=\,$4$\,$K and $B$$\,=\,$0$\,$T [Fig. 1b]. $\rhoxx$ drops to 20$\,$$\Omega$ at n=5x10$^{12}$cm$^{-2}$, although in this  regime the mean free path is limited by boundary scattering~\cite{Wang2014} and the sheet resistivity is not well defined. The quantum mobility is also estimated from the onset of the $\nu$$\,=\,$2 gaps in $\rhoxx(\Vbg,B)$~\cite{Bolotin2008}, which for example occurs at 25$\,$mT for Device B [Fig. 1b Inset], indicating a mobility of at least 4$\times$$10^{5}$ cm$^{2}$/Vs. 

Five of our seven devices are strongly insulating at the charge neutrality point, with peak resistivities $\rho_{NP}$ exceeding 100$\,$k$\Omega$ at 4$\,$K, well above the theoretical limit of $\pi h/4e^{2}$$\sim$20$\,$k$\Omega$ expected for gapless pristine graphene~\cite{Nilsson2006,Tworzydlo2006,Gorbar2002} (the last two devices are comparable in quality but with $\rho_{NP}<$20$\,$k$\Omega$). This phenomenon has only been seen in extremely clean graphene on boron nitride~\cite{Hunt2013,Amet2013} and has been attributed to sublattice symmetry breaking~\cite{Giovannetti2007,Levitov2013,Kindermann2012,Hunt2013,Amet2013}. We made no attempt to rotationally align our graphene and BN flakes, suggesting that very close rotational alignment is not required to produce this insulating behavior (in agreement with Ref.~[\citen{Hunt2013}] where a graphene flake with misalignment up to 4 degrees on BN still showed activated transport). In only 2 of our 5 insulating devices do we see superlattice peaks at attainable densities (5.3 and 6.0$\times 10^{12}$cm$^{-2}$ respectively), indicating rotational misalignment under 1.5 degrees [See supplementary figures]. In our most insulating device [Device A], $\rho_{xx}(\Vbg)$ spans 6 orders of magnitude, up to 20$\,$M$\Omega$ at the neutrality point [Fig. 1b, in log scale]. The temperature dependence of its peak resistivity is activated above 20$\,$K [Fig. 1b]: an Arrhenius fit $\rhoxx$$\sim$exp(-$\Delta/2\kB T$) yields a gap $\Delta$$\,=\,$350$\,\pm\,$60$\,$K, similar to that reported recently~\cite{Hunt2013}. 
  
As $B$ is increased, vanishing $\rhoxx$ indicates quantum Hall phases at integer filling factor [dark blue regions on Fig. 1c]. Broken symmetry states $\nu$$\,=\,$1 and 3 are visible at $B\sim$1.5$\,$T, and the zero field insulating state continuously undergoes a transition to the $\nu$$\,=\,$0 phase below 1T~\cite{Amet2013, Hunt2013}. Remarkably, FQH states at $\nu$$\,=\,$-8/3, -10/3, -11/3 are seen for $B$ as low as 5$\,$T, significantly lower than in Ref.~[\citen{Dean2011,Du2009,Bolotin2009}]. From now on we focus on FQH states in the hole doped regime. Contrary to the case of bilayer graphene~\cite{Kou2014}, the sequence of FQH states in monolayer graphene is electron-hole symmetric, but our data are significantly cleaner on the hole side. We suppress the negative sign of $\nu$ when we refer to values of filling factor. 

\subparagraph{Fractional Quantum Hall Effect in the Zeroth Landau level}
$\\$

The zeroth Landau level in monolayer graphene is populated at filling factors $\vert \nu\vert\leq2$. Between $\nu$$\,=\,$0 and 1, we observe a detailed series of FQH states following the two-flux composite-fermion sequence~\cite{Feldman2012} at $\nu$ (or $1-\nu$)$\,=\,$1/3, 2/5, 3/7, 4/9 [Fig. 2a]. Corresponding plateaus appear in the transverse conductivity $\sigmaxy$$\,\equiv\,$$\rhoxy/(\rhoxx^{2}+\rhoxy^{2})$, despite modest mixing between longitudinal and transverse signals.

The sequence of FQH states between $\nu$$\,=\,$1 and 2 is more intriguing: transport and local compressibility measurements~\cite{Dean2011, Feldman2012,Feldman2013} in graphene both suspended and on boron-nitride have lacked incompressible states at odd-numerator fractional filling factors. The associated ground-states are believed to be spin-polarized, but susceptible to valley-textured excitations ~\cite{Abanin2013}. Unlike more familiar spin skyrmions~\cite{Sondhi1993}, spin-polarized valley skyrmions can be spatially extended even at high magnetic fields~\cite{Abanin2013} because they do not involve spin-flips. They provide low-lying delocalized excitations and explain the absence of incompressible states at these odd numerator fractions in Ref.~[\citen{Dean2011, Feldman2012}]. Yet in Figure 2b we see minima of $\rhoxx$ not only at even numerator filling factors 4/3 and 8/5 but also at 5/3 and 7/5. These were absent in Ref.~[\citen{Dean2011, Feldman2012}] and suggest that valley skyrmions are suppressed in our samples.

We extract activation gaps from Arrhenius fits to the temperature dependence of $\rhoxx$ at 25T [Fig. 2c]. Surprisingly, we find $\Delta_{5/3}$ to be the most robust at 10.4$\Pm$1.4$\,$K, followed by  $\Delta_{4/3}$\=8.9$\Pm$0.6$\,$K, $\Delta_{2/3}$\=4.6$\Pm$.3$\,$K and $\Delta_{1/3}$\=3.5$\Pm$.2$\,$K. Theoretical estimates of these gaps are one order of magnitude larger. $\Delta_{4/3}$ and $\Delta_{2/3}$ were predicted to be the largest gaps, ranging from 0.08 to 0.11 e$^{2}$/$\epsilon\lB$ (or 21-29$\,$K$\times$$\sqrt{B[T]}$)$\,$, for $\epsilon$$\,=\,$$(1+\epsilon_{BN})/2\approx2.5$)~\cite{Shibata2009,Toke2012}. Theoretical estimates for $\Delta_{1/3}$ and $\Delta_{5/3}$ are slightly smaller: 0.03 to 0.1 e$^{2}$/$\epsilon\lB$ (or 8-26$\,$K$\times$$\sqrt{B[T]}$$\,$)~\cite{Toke2012,Alpakov2006}.  The observed gaps are most likely reduced relative to predictions due to disorder, the density at which FQH states occur fluctuating spatially due to remaining charged impurities ~\cite{Feldman2012}. This sensitivity to disorder causes FQH gaps to sometimes decrease after successive cool-downs, and to vary slightly between samples. It is also possible that gaps are reduced as filling factors approach $\nu$$\,=\,$0 due to the competing insulating phase at charge neutrality~\cite{Dean2011, Feldman2012}.

The boron nitride substrate can dramatically affect the band structure of graphene due to the very similar crystal lattices of these two materials~\cite{Yankowitz2012}. Satellite Dirac peaks emerge at high densities~\cite{Ponomarenko2013, Dean2012, Hunt2013} and the quantum Hall spectrum changes dramatically, exhibiting a  fractal pattern known as the Hofstadter butterfly when $l_{B}$ is comparable to the superlattice moiré cell size~\cite{Ponomarenko2013, Dean2012, Hunt2013}. We observed superlattice effects in some of our devices at very high densities~\cite{Suppinfo}, but restricted our study to samples and ranges of ($n,B$) where such effects do not impact the quantum Hall spectrum. Even for twist angles up to a few degrees, boron nitride may break the sublattice symmetry~\cite{Hunt2013, Levitov2013}, resulting in a zero-field insulating behavior at the neutrality point as observed in Fig. 1b~\cite{Hunt2013, Amet2013}.  In the zeroth Landau level, wavefunctions corresponding to the two valleys are each localized on a different sublattice. Sublattice symmetry breaking should therefore yield valley-polarized ground states, making extended valley skyrmions energetically costly. This is consistent with the observation of FQH states at $\nu$$\,=\,$5/3~\cite{Hunt2013} and 7/5 in samples that are also gapped at zero field. 

\subparagraph{Fractional Quantum Hall Effect in the First Landau level, $\nu$$\,=\,$2 to 6}
$\\$

FQH states in the first LL of graphene were predicted to be much more robust than in non-relativistic 2DEGs~\cite{Alpakov2006, Toke2006}, yet only states with denominator 1/3 have been reported in transport, and scanning compressibility measurements have not reached sufficient density to probe beyond the zeroth LL. In contrast, we observe a detailed sequence of FQH states in transport, with sharp local minima of $\rhoxx$ visible at factors $\nu+\delta\nu$, where $\nu$$\,=\,$2, 3, 4 and $\delta\nu$$\,=\,$1/3, 2/3, 2/5, 3/5, 3/7, 4/7... [Fig. 3a, Device A, $T$$\,=\,$300$\,$mK]. Minima of $\rhoxx$ occur at constant filling factors (vertical blue lines) indicating that they are indeed FQH states, some of them emerging at particularly low fields. Between $\nu$$\,=\,$2 and 3, additional FQH states are noticeable at 45$\,$T, such as 17/7, 22/9 and 27/11 [Fig. 3f]. We do not observe fractions beyond $\nu$$\,=\,$5 and even the integer QH state $\nu$$\,=\,$5 is not always clearly defined, a feature we have observed in several devices and cannot presently explain.

As expected, $\rhoxx$ minima have a strong temperature dependence, captured at 14T for a different device (B) in Figure 3b-d for temperatures between 500$\,$mK and 4$\,$K. For each filling factor, $\rhoxx(T)$ is fitted to an Arrhenius law $\rhoxx(\nu)\sim$ exp(-$\Delta_{\nu}/2\kB T)$. At that field, we find $\Delta_{8/3}$\=5.5$\Pm$1.5$\,$K and $\Delta_{10/3}$\=6.3$\Pm$2.1$\,$K, while $\Delta_{7/3}$\=5.4$\Pm$1.4$\,$K, $\Delta_{13/3}$\=1.2$\Pm$.2$\,$K and $\Delta_{14/3}$\=2.3$\Pm$.5$\,$K. The temperature dependence of $\rhoxx(11/3)$ is not strong enough to extract a gap value. As expected, FQH states with higher denominator are weaker than denominator 3 states at nearby filling factors, with $\Delta_{12/5}$\=1.4$\Pm$.3$\,$K, $\Delta_{13/5}$\=2.6$\Pm$.8$\,$K and $\Delta_{17/5}$\=1.8$\Pm$.4$\,$K.  

\subparagraph{Probing the Nature of the Fractional Quantum Hall States}
$\\$

The extracted gaps are at least one order of magnitude smaller than theoretically predicted, and are likely reduced by disorder. In addition, their perpendicular field ($B_{\perp}$) dependence does not follow that of the Coulomb energy at the magnetic length scale $\Ec$$\,\propto\,$$\lB^{-1}$$\,\propto\,$$\sqrt{B_{\perp}}$. Rather, the field dependence we observe is not monotonic in the first LL, with a noticeable decrease at high fields we did not observe in the zeroth LL [Device A, Fig. 3g]. This could be due to emerging superlattice effects at high field and density, or to approaching fractional quantum Hall phase transitions between different ground states. 
 
To probe the spin polarization of the ground states at each filling factor, we follow the evolution of the gap when the sample is tilted with respect to $B$, which allows us to change the relative strength of the Zeeman coupling, proportional to the total field $B$, and Coulomb interactions, controlled by the perpendicular field $\Bz$ [Fig. 4 Inset].  In what follows, we call the in-plane field $B_{\parallel}$, so $B=\sqrt{B_{\parallel}^{2}+B_{\perp}^{2}}$. Previous graphene transport measurements in the FQH regime required very large $\Bz$, making it challenging to tilt samples with respect to the magnetic field while preserving FQH states. Here, fractions are observed at fields as low as 6$\,$T, so we can measure $\rhoxx$ at a fixed $\Bz\,=\,$17$\,$T while substantially changing the total field [Fig. 4]. In the zeroth Landau level, the minima of $\rhoxx$ at $\nu\,=\,$1/3, 2/5, 3/5, 2/3, 4/3 and 5/3 do not change strength when increasing $B_{\parallel}$ with $\Bz$ fixed, consistent with spin polarized states. This is in agreement with Ref.~[\citen{Feldman2013}], where phase transitions to a presumed spin-polarized state were observed at every filling factor other than 1/3 in the zeroth LL, with transition $\Bz$ increasing with the composite fermion LL index $p$ and ranging from 6 to 12$\,$T, lower than the field range studied here. 

This contrasts with our observations in the first Landau level, where most FQH gaps are seen to decrease with $B_{\parallel}$. Minima at 17/5 and 18/5 disappear, while 10/3 and 11/3 are visibly weakened. Only 8/3 remains robust. We observed the same behavior in two different samples at perpendicular field up to 25$\,$T. We would naively expect that such a high field should be strong enough to completely polarize the spin, but gaps weakened by increasing $B_{\parallel}$ suggest instead that the first LL FQH ground states are not yet spin-polarized even at 25$\,$T. Integer quantum Hall gaps follow a similar evolution than the one observed in Ref.~[\citen{Young2012}]: at half-filling, the region of vanishing $\rhoxx$ widens with $B_{\parallel}$ suggesting a strengthening of $\Delta_{4}$, while gaps at quarter-filling $\Delta_{3}$ and $\Delta_{5}$ remain the same. This was attributed in Ref.~[\citen{Young2012}] to a spin-polarized ground-state at $\nu$$\,=\,$4 with low-lying excitations involving spin-flips, while ground states at quarter filling have valley-textured excitations. 

\paragraph{Discussion}
$\\$

The spin polarization of the ground state at a given filling factor depends on the competition between electron interactions and the Zeeman splitting. Neglecting spin, FQH gaps between two successive integer filling factors are expected to scale like Coulomb interactions, with $\Delta_{\nu}\propto \vert1-2\delta\nu\vert e^{2}/\epsilon\lB$, where $\delta\nu=p/(2p+1)$. The prefactor $\vert 1-2\delta\nu\vert$ stems from the effective field $B^{*}=B_{\perp}\vert1-2\delta\nu\vert$ experienced by composite fermions, which vanishes when approaching half filling. This explains the decreasing gaps for FQH states with high index $p$, a well-documented fact in GaAs-based 2DEGs~\cite{Du1993, Du1995}. 

For CFs with a spin, the spin polarization of the ground state at a given filling factor depends on the competition between electron interactions and the Zeeman splitting: Zeeman splitting of two consecutive CF Landau levels should produce a LL crossing when $g\mu B$ equals the Coulomb contribution to the gap stated above, which only scales like $\sqrt{B_{\perp}}$. Below the transition, activation gaps should decrease linearly with an in-plane field, which could explain our observations in the first LL. Such crossings and associated transitions of ground states have been observed in transport in GaAs-based 2DEGs~\cite{Du1995}, and more recently in the zeroth Landau level of graphene using compressibility measurements~\cite{Feldman2013}. 

How should field scales for these transitions differ between GaAs and graphene? The dielectric constant is 5 times lower in graphene on BN than in GaAs, so at a given field electron interactions are stronger in graphene. Yet, Zeeman splitting is also much larger in graphene, with $g$$\,\geq\,$2 compared to $g$$\,\approx\,$0.4 in GaAs, and it is therefore unclear why the field required to observe transitions to spin-polarized states should be larger in graphene. However for a given ratio $e^{2}/\epsilon\lB$ FQH gaps in the first Landau level of graphene were predicted to be significantly stronger than gaps in both non-relativistic 2DEGs ~\cite{Alpakov2006,Toke2006} and graphene's zeroth LL, a result of the peculiar form factor of relativistic electron in $n\neq0$ Landau levels. We could not quantitatively compare gaps between zeroth and first Landau levels at the same field in our devices~\cite{Suppinfo}. However at low field ($B$$\,\leq\,$14T), several devices showed better developed FQH states in the first Landau level~\cite{Suppinfo}, in qualitative agreement with theoretical expectations~\cite{Alpakov2006,Toke2006}. This could explain the need for larger Zeeman splitting to overcome interaction-induced gaps in the first Landau level and induce phase transitions to spin-polarized states, but would not explain what is special about $\nu$$\,=\,$8/3, the only fractional state in the first Landau level which is not weakened by an in-plane field at $B_{\perp}$$\,=\,$17T.

\paragraph{Methods}
$\\$
The graphene/h-BN/graphite structure is fabricated using the transfer method described infit Ref.~[\citen{Amet2013,Amet2014}]. Few-layer graphite is exfoliated on SiO$_{2}$. Graphene and h-BN flakes are separately exfoliated on a polymer stack of polymethyl-methacrylate (400$\,$nm) atop polyvinyl alcohol (60$\,$nm). We transfer h-BN first, then graphene, on top of the graphite flake. For each transfer, the PVA is dissolved in deionized water, the PMMA membrane lifts-off, is transferred on a glass slide, heated at 110$^{\circ}$C and aligned on top of the target with a micro manipulator arm. After each transfer, samples are annealed in flowing argon and oxygen (500 sccm and 50 sccm respectively) in a 1" tube furnace at 500$^{\circ}$C to remove organic contamination~\cite{Garcia2012}. E-beam lithography is combined with oxygen plasma etching to define the graphene flake geometry, and with thermal evaporation to deposit Cr/Au contacts (1nm/100nm). 

The devices were measured in a $^{3}$He cryostat and a dilution fridge in a current-biased lock-in setup, with an ac excitation current of 20$\,$nA at 13$\,$Hz. Measurements at fields higher than 14$\,$T were performed at the National High Magnetic Field Laboratory in Tallahassee.

\paragraph{Acknowledgments}
$\\$
We thank Mark Goerbig, Dmitri Abanin, Steve Kivelson and Chi-Te Liang for their input. We are grateful for Jonathan Billings' technical help. Part of the measurements were done at the National High Magnetic Field Laboratory, which is supported by US National Science Foundation cooperative agreement no.DMR-0654118, the State of Florida and the US Department of Energy. F. A. is funded by the Center for Probing the Nanoscale, an NSF NSEC, supported under grant No. PHY-0830228. A.J.B. is supported by a Benchmark Stanford Graduate Fellowship.

\clearpage
\paragraph{References}

\clearpage
              							%%%%%%% PUT THE FIGURES HERE %%%%%%%%%%
\paragraph{Figure legends}
\subsection{Figure 1}
Characterization at low-field. (a) Schematic of the device, consisting of a graphene/h-BN/graphite stack resting on SiO$_{2}$. $\Vbg$ is applied to the graphite back gate to tune the carrier density. (b) $\rhoxx(n)$ in logarithmic scale, measured at $B$$\,=\,$0$\,$T and $T$$\,=\,$4$\,$K. Inset: $\rhoxx(n,B)$ at very low field and density. (c) Fan diagram of $\rhoxx(n,B)$ up to 11$\,$T.

\subsection{Figure 2}
Fractional Quantum Hall effect in the zeroth Landau Level. (a) $\rhoxx(\nu)$ between $\nu$$\,=\,$0 and $\nu$$\,=\,$1 at 25$\,$T (red), 35$\,$T (purple) and 45$\,$T (blue). At 45$\,$T FQH states with denominator 9 are noticeable. $\sigma_{xy}$ (black curve) shows well-defined plateaus at 2/3, 3/5 and 4/7. At lower filling factors, mixing with $\rhoxx$ makes the plateaus indistinct. (b) $\rhoxx(\nu)$ between $\nu$$\,=\,$1 and $\nu$$\,=\,$2 for $B$ ranging from 17$\,$T to 25$\,$T. FQH states are visible at 5/3 and 7/5, not just even-numerator fractions. $\sigma_{xy}$ (black curve) at 17$\,$T. (c) Arrhenius plots for $\nu$$\,=\,$1/3, 2/3, 4/3 and 5/3, showing activated behavior. 

\subsection{Figure 3}
Fractional Quantum Hall Effect in the first Landau Level. (a) $\rhoxx(n,B)$ from $\nu$$\,=\,$2 to $\nu$$\,=\,$5, measured at B=14$\,$T, T=30$\,$mK [Device B]. FQH states follow the composite fermion sequence p/(2p$\pm$1) between each integer up to $\nu$$\,=\,$5. 7/3 is poorly resolved in this device, but this is not a generic feature. No FQH states are seen beyond $\nu$$\,=\,$5. (b-d) Temperature dependence of the FQH states in the $\nu$$\,=\,$2 to 3, $\nu$$\,=\,$3 to 4 and $\nu$$\,=\,$4 to 5 sequences respectively [Device A]. Temperatures range from 500$\,$mK to 4$\,$K. $\rhoxx$ exponentially vanishes as $T$ is lowered. Contrary to Device B, the 7/3 state is visible in this device. (e) $\rhoxx$ and $\sigma_{xy}$ in the $\nu$$\,=\,$2 to 3 sequence, measured at B=14$\,$T and T=30$\,$mK [Device A]. Plateaus of $\sigmaxy$ are seen at filling factors 7/3, 12/5, 13/5, 8/3. FQH states are observed in $\rhoxx$ at $\nu$$\,=\,$17/7 and 18/7 but are not resolved in $\sigmaxy$. (f) $\rhoxx$ observed in the $\nu$$\,=\,$2 to $\nu$$\,=\,$3 sequence (45$\,$T, T$\,=\,$400$\,$mK, Device A). FQH states are observed at denominators up to 11. (g) Field dependence of the activation gaps for Device B at $\nu$$\,=\,$8/3 (red), 11/3 (black) and 13/3 (yellow).

\subsection{Figure 4}
In-plane field dependence of the FQH states in the first two Landau levels of Device A at T$\,=\,$400$\,$$\,$mK. (a) $\rhoxx(\nu)$ between $\nu$$\,=\,$0 and $\nu$$\,=\,$2 ($\Bz$$\,=\,$17$\,$T, $B$$\,=\,$17$\,$T and 44T). (b) $\rhoxx(\nu)$ between $\nu$$\,=\,$2 and $\nu$$\,=\,$5 ($\Bz$$\,=\,$17$\,$T, $B$$\,=\,$17$\,$T, 33$\,$T and 44T). Inset: Schematic of the in-plane field dependence of the composite fermion LL energies at fixed $B_{\perp}$. (c) $\rhoxx(\nu)$ between $\nu$$\,=\,$3 and $\nu$$\,=\,$4 ($\Bz$$\,=\,$25$\,$T, $B$$\,=\,$25$\,$T and 45$\,$T).
\clearpage

\begin{figure*}[t!]
\center \label{fig1}
\includegraphics{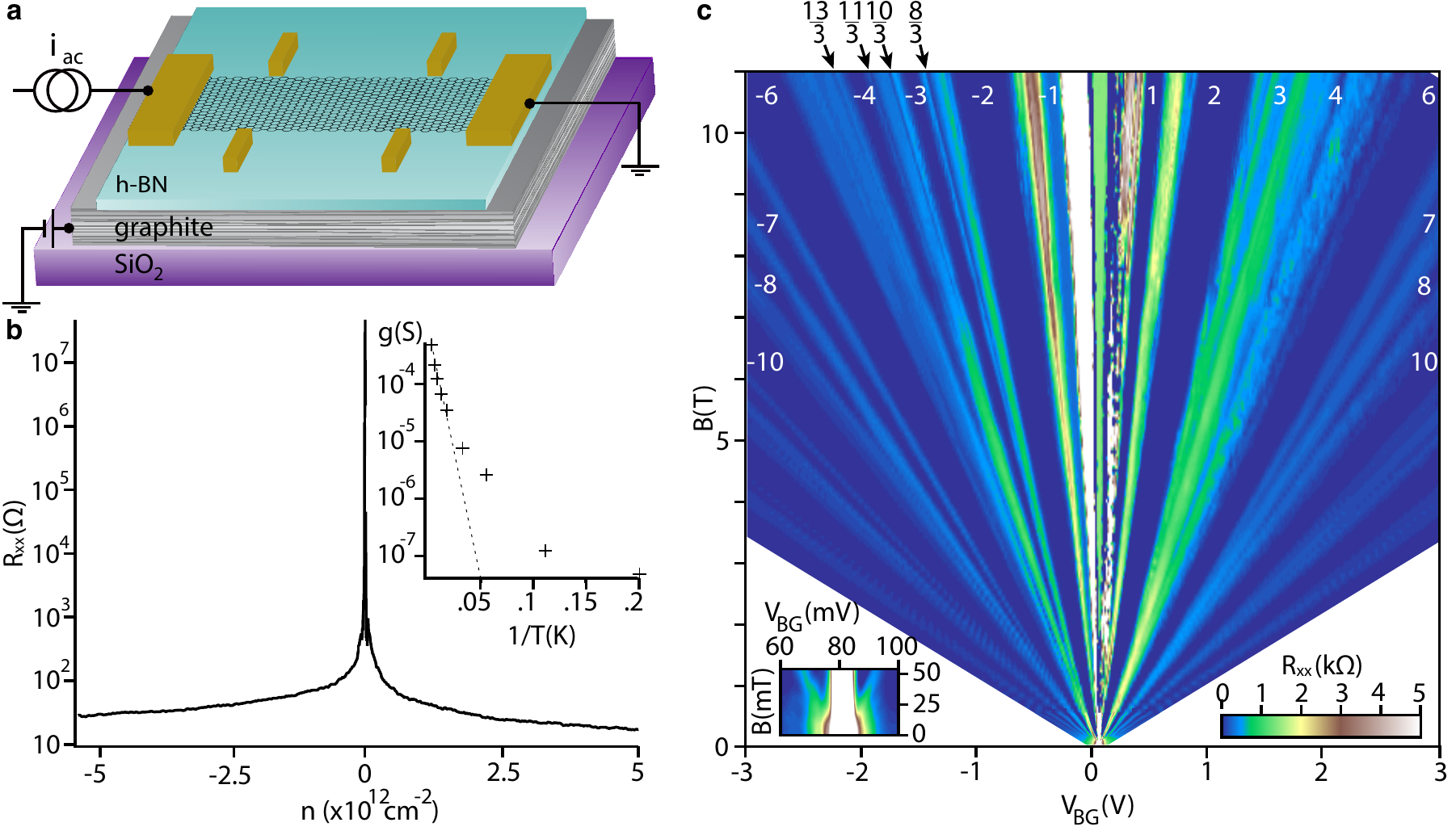}
\caption{}
\end{figure*}

\begin{figure*}[t!]
\center \label{fig2}
\includegraphics[width=180mm]{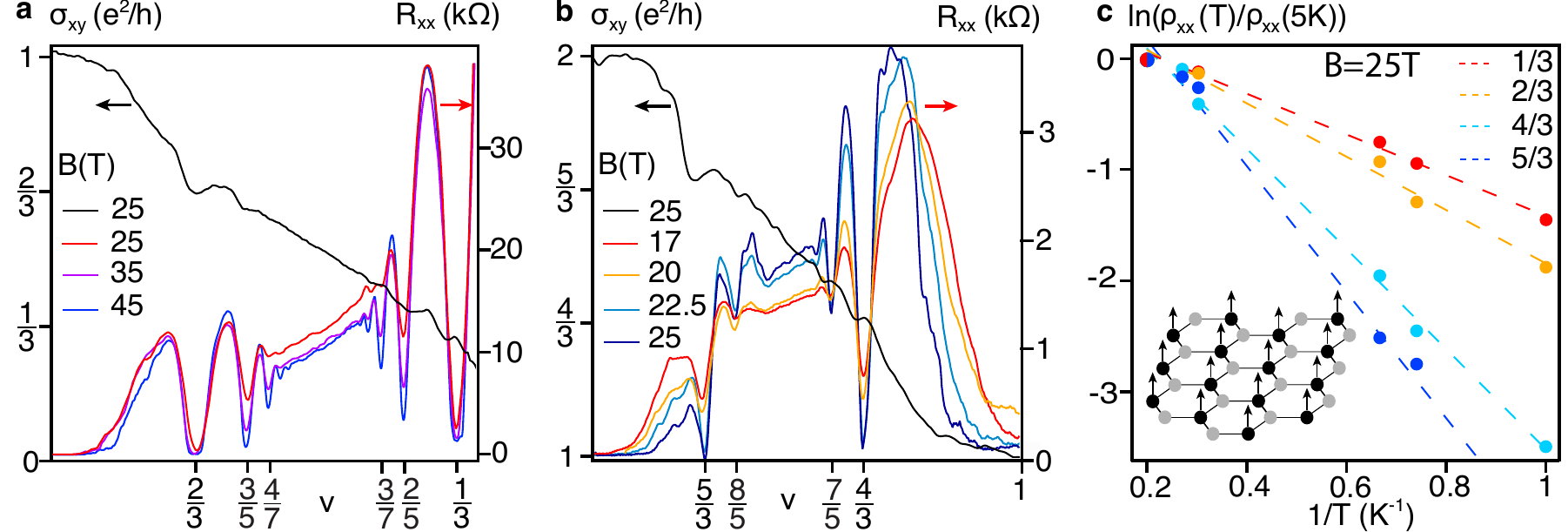}
\caption{}
\end{figure*}

\begin{figure*}[t!]
\center \label{fig3}
\includegraphics[width=180mm]{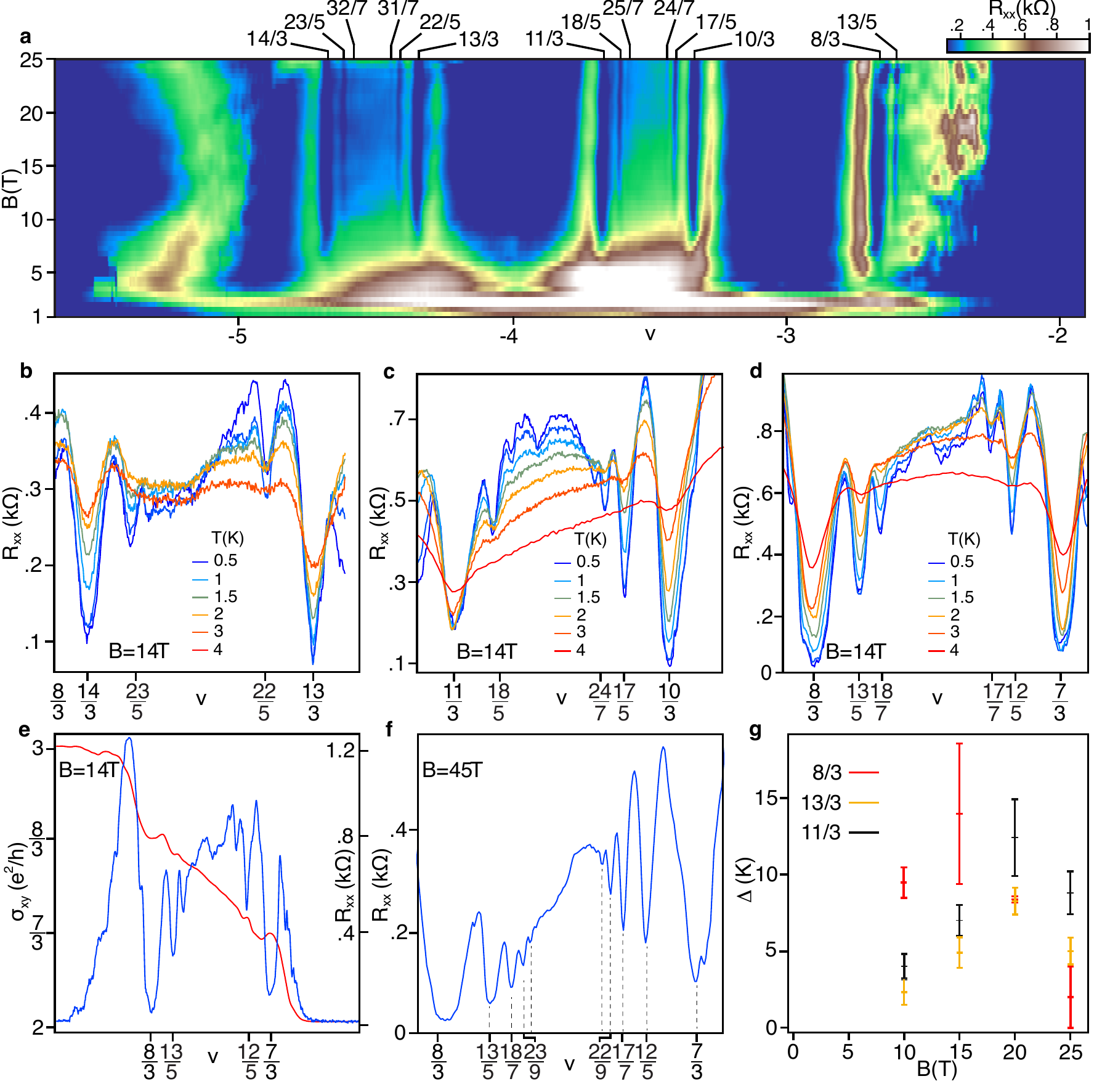}
\caption{}
\end{figure*}

\begin{figure*}[t!]
\center \label{fig4}
\includegraphics[width=180mm]{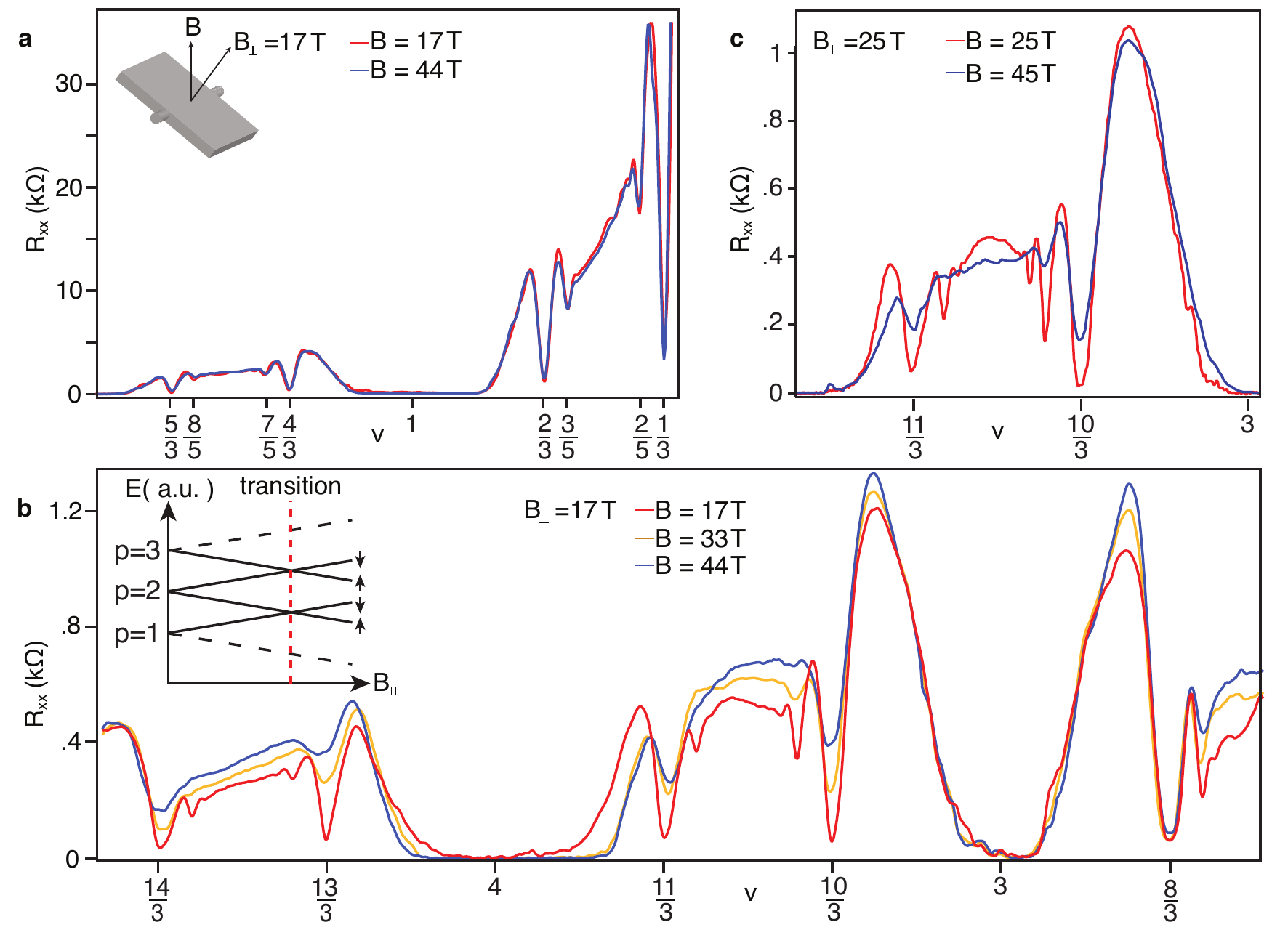}
\caption{}
\end{figure*}

\clearpage

{\centering\chapter{\textbf{Supplementary information}}}

\subsection{Insulating behavior and superlattice effects at zero field} 

We observed very large peak resistivities at the neutrality point in several high-mobility devices, as shown for three of them on Figure 5.  These are measured at zero field and T$\,=\,$400$\,$mK. In addition to being gapped, Devices C and D show satellite Dirac peaks characteristic of superlattice effects when the graphene and boron nitride crystalline orientation are closely aligned. These satellite peaks are observed at densities n$\,=\,$$6.0\times10^{12}$cm$^{-2}$  and $\,=\,$$5.3\times10^{12}$cm$^{-2}$ for devices C and D respectively, which corresponds to a misalignment of 1.3$^{\circ}$ and 1.1$^{\circ}$. The fractional quantum Hall spectrum in these two devices was altered by Hofstadter butterfly physics~\cite{Hunt2013, Ponomarenko2013, Dean2012}, which is not the purpose of the present paper.

\subsection{In-plane field dependence of the FQH states in the first Landau level}

We show on Figure 6 the longitudinal resistance of Device B in the first Landau level, measured at T$\,=\,$400mK. The perpendicular field $B_{\perp}$ is kept  at 10T while the total field $B$ is varied. Integer states at quarter filling $\nu$$\,=\,$3 and 5 don't change noticeably while $\nu$$\,=\,$4 is visibly stronger. Similar to what was observed for device A, fractional quantum Hall states are all weakened by an in-plane field except for $\nu$$\,=\,$8/3. No FQH states were observed at 7/3 and beyond $\nu$$\,=\,$5. 

\subsection{Relative strengths of gaps in the zeroth and first Landau level}

We find that observation of crisp FQH states in the zeroth LL often required higher field than in the first LL. We show on Figure 7 the longitudinal resistance in four devices at relatively low fields (10T and 14T). FQH states are generally better developed in the first LL than in the zeroth LL. Devices E and F don't show any FQH states in the zeroth LL [Fig. 7(a-b)]. In Device B, zeroth LL FQH states start being visible at 14T but are not as well developed as in the first LL, where denominators up to 7 are observed, along with nascent four-flux composite fermion states at 11/5 and 14/5 [Fig. 7(c), blue arrows].  Figure 7(d) shows an exception, Device C, where FQH states are already well developed in the zeroth L (including at 5/3), and comparable to the first LL states. 

At much higher field, however, we observed a non monotonic field dependence of the FQH gaps in the first LL leading to their degradation. This difference, along with device contamination after successive cool-downs, prevented us from tracking the field dependence of FQH gaps in both levels over a large range of fields. 

\newpage

\paragraph{Supplementary figures legends}

\subsection{Figure 5}
(a) Device B [in log scale], with a peak resistivity of 90$\,$k$\Omega$ and a mobility of 4$\times 10^{5}$ cm$^{2}/Vs$ at low density. Fractional quantum Hall data from Device A 400nm are shown in the main part of the paper. (b) Device C [log scale], with a peak resistivity of 1$\,$M$\Omega$ and a mobility of approximately $10^{6}$ cm$^{2}/Vs$ at low density. (c) Device D, with a minimum conductivity of 1 $\upmu$S measured in a two-terminal geometry [inset in log scale]. The mobility is approximately $1.5\times10^{5}$ cm$^{2}/Vs$ at low density. The corresponding peak resistivity is 1$\,$M$\Omega$.

\subsection{Figure 6}
In-plane field dependence of the FQH states in the first Landau level of Device B at T$\,=\,$400$\,$mK. $\rhoxx(\nu)$ is measured between $\nu$$\,=\,$2 and $\nu$$\,=\,$5 ($\Bz$$\,=\,$10$\,$T, $\Bt$$\,=\,$10$\,$T, 22$\,$T and 31T). All FQH states ---except 8/3--- are weakened by an in-plane field.

\subsection{Figure 7}
(a) $\rhoxx(V_{BG})$ for device E measured between $\nu$$\,=\,$0 and $\nu$$\,=\,$6 at T$\,=\,$30mK and $B$$\,=\,$10$\,$T. Nascent FQH states are indicated with red arrows. (b) $\rhoxx(V_{BG})$ for device F measured between $\nu$$\,=\,$0 and $\nu$$\,=\,$6 at T$\,=\,$30mK and $B$$\,=\,$11$\,$T. $\rhoxx$ is scaled down by a factor 1/15 below $\nu$$\,=\,$2. (c) $\rhoxx(V_{BG})$ for device B measured between $\nu$$\,=\,$0 and $\nu$$\,=\,$6 at T$\,=\,$30mK and $B$$\,=\,$14$\,$T. $\rhoxx$ is scaled down by a factor 1/20 below $\nu$$\,=\,$2. The sequence FQH states is already very well developed up to denominator 7. Possible four-flux CF states are indicated with blue arrows at $\nu$$\,=\,$11/5 and 14/5. (d) $\rhoxx(V_{BG})$ for device B measured between $\nu$$\,=\,$0 and $\nu$$\,=\,$6 at T$\,=\,$300mK and $B$$\,=\,$14$\,$T. $\rhoxx$ is scaled down by a factor 1/10 below $\nu$$\,=\,$2.

\newpage
\begin{figure*}[t!]
\center \label{fig5}
\centerline{\includegraphics{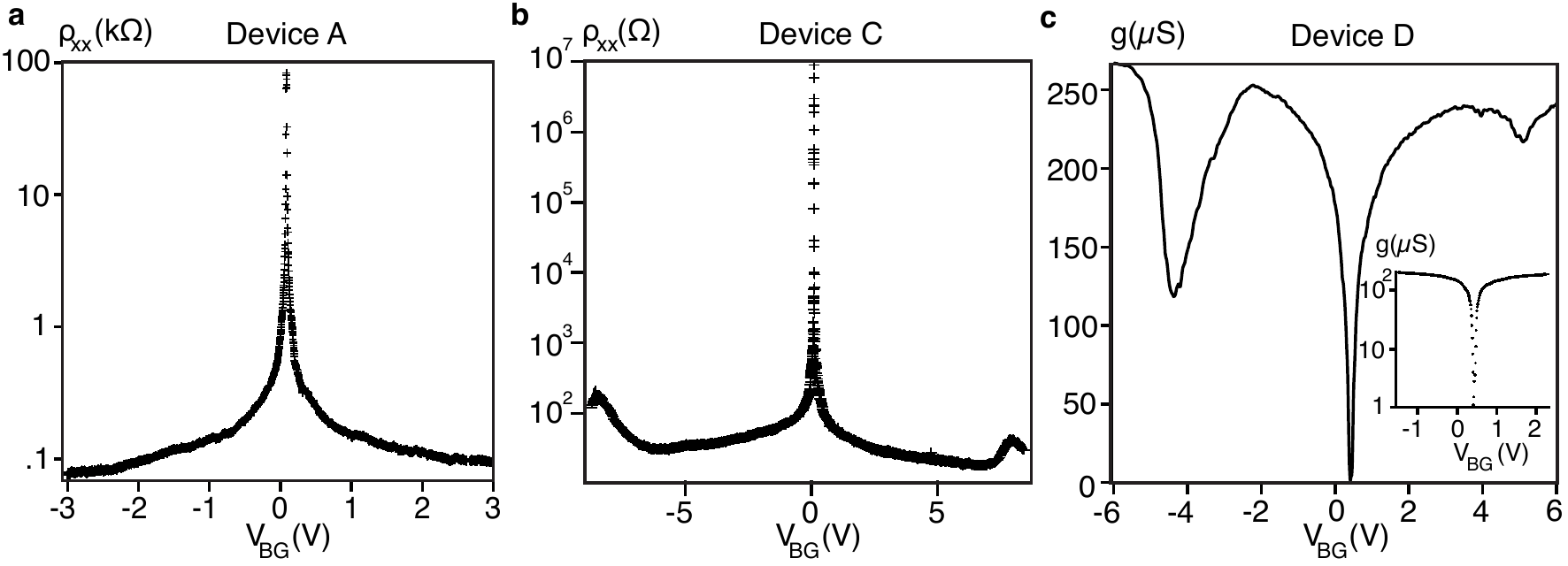}}
\caption{}
\end{figure*}

\begin{figure*}[t!]
\center \label{fig6}
\centerline{\includegraphics{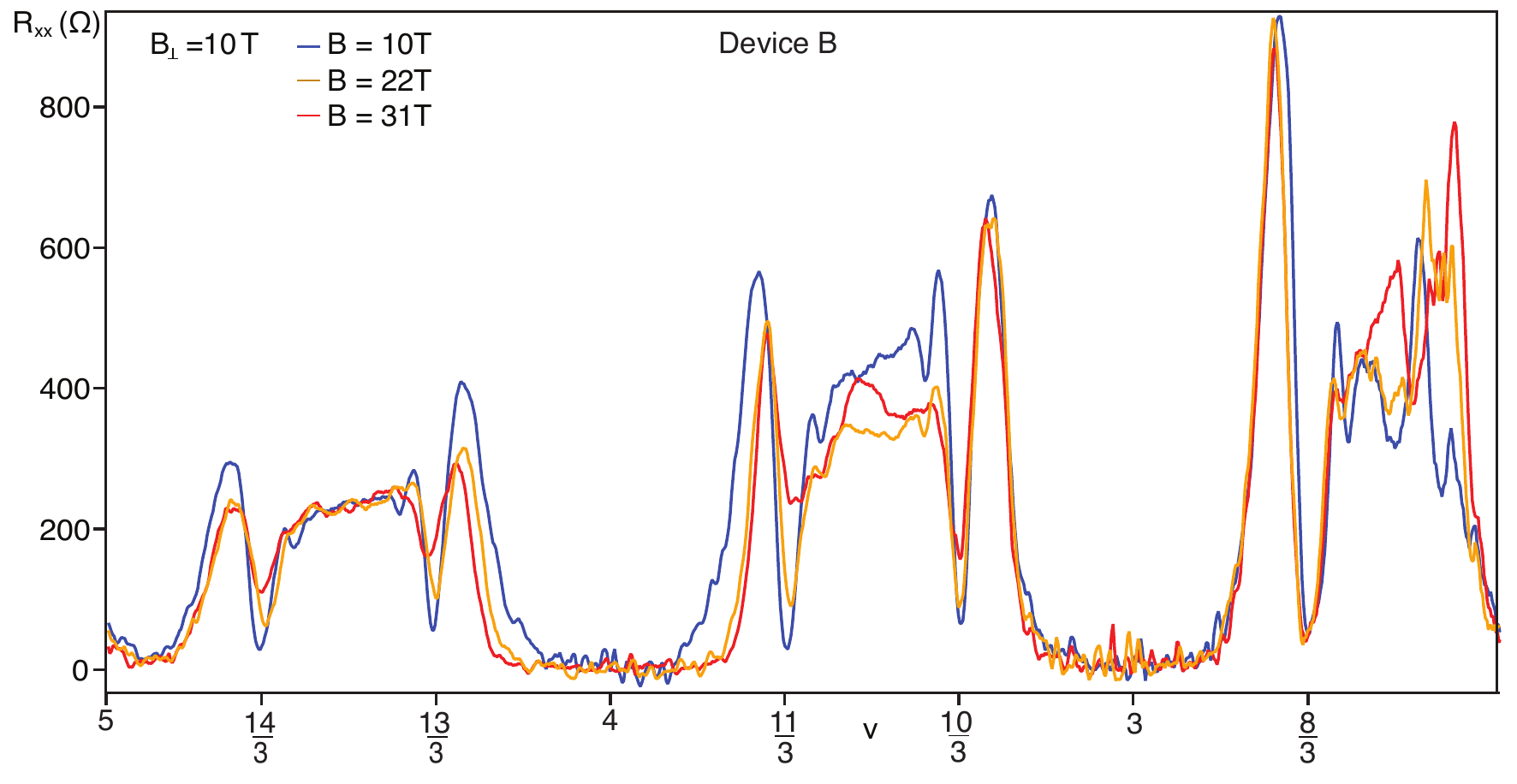}}
\caption{}
\end{figure*}

\begin{figure*}[t!]
\center \label{fig7}
\centerline{\includegraphics{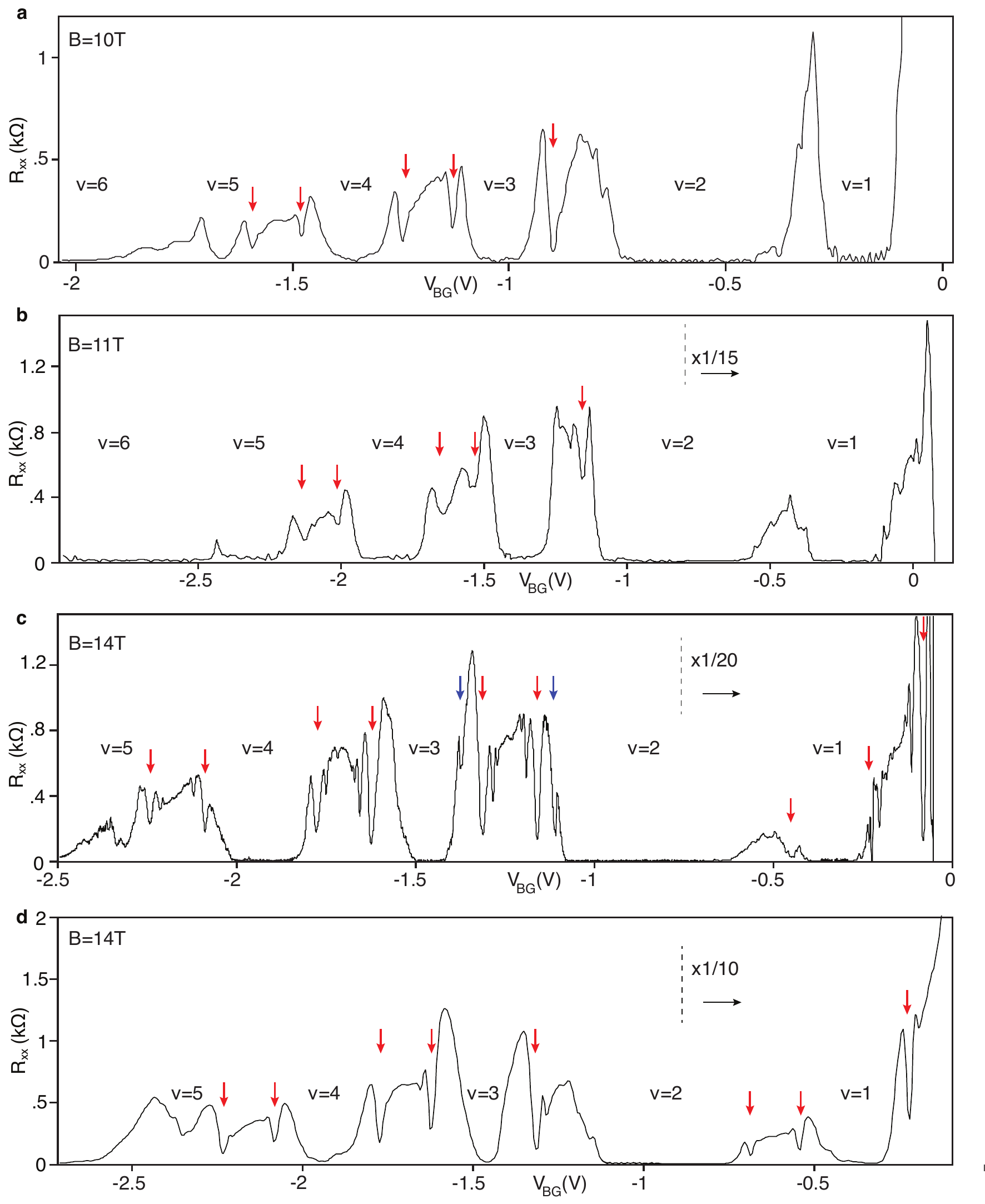}}
\caption{}
\end{figure*}

\end{document}